\documentstyle[psfig,11pt]{article}
\def\asca{{\it ASCA}}

\def\rosat{{\it ROSAT}}

\def\myarcmin{^\prime\mskip-5mu}

\begin{document}
\vspace{1.0cm}
{\Large \bf \asca\ OBSERVATIONS OF TWO SNRS AND NEI ANALYSIS}

\vspace{1.0cm}

Sun M. , and Wang Z.R.

\vspace{1.0cm}
{\it Department of Astronomy, Nanjing University, Nanjing, 210093, P.R.China\\
Email: quqiny@nju.edu.cn ; zrwang@nju.edu.cn}
\vspace{0.5cm}
\section*{ABSTRACT}
Based on the data from the \asca\ observation of SNRs Kes79 and W49B, we present here the
analysis of their X-ray spectra and morphologies. The Kes79 spectrum
can be well fitted by a single NEI component, and the narrow-band images of that source 
show an inhomogeneous distribution of heavy elements. The heavy elements are richest in
the positions S, SE and SW of Kes79, where there may exist interaction between shocks and
molecular clouds implied by radio observations. For W49B we present here the non-equilibrium
ionization (NEI) analysis
based on its emission line diagnostics, and the spectral fit using two NEI components. The reverse
shock in W49B may be still hot and we don't find evidence for a hotter blast wave in \asca\
spectra.

\section{INTRODUCTION}
In radio Kes79 is a shell type SNR but with a complex structure. It has an incomplete outer 
shell with radius $\sim$ 6 $\myarcmin\mbox{ }$ and two prominent indentations in the east and
northeast. And it also has strong emission from the center - "an inner ring" (Frail \& Clifton, 1989). No
pulsar has been found (Velusamy et al. 1991). Scoville et al. (1987), Green \& Dewdney (1992; hereafter GD)
found strong CO emission from the east, southeast and south of this source. And GD also found
bright, extended HCO$^{+}$ emission from the east of this source. GD estimated the global average
density of the molecular cloud surrounding Kes79 at $\sim$ 180 cm$^{-3}$ and showed that appreciable 
HCO$^{+}$ emission requires a density of 10$^{5}$ cm$^{-3}$. Green et al. (1997)
detected OH (1720 MHz) maser emission from this source. They suggested that this emission is
due to the interaction between shock waves and molecular materials. In X-ray Kes79 has strong
central emission and diffuse outer emission with diameter $\sim$ 5 $\myarcmin\mbox{ }$. Seward \&
Velusamy (1995) once used the Sedov model to analyse this source according to the \rosat\ data.
Its distance is about 10$\pm$2 kpc (Frail \& Clifton, 1989) based on 21 cm neutral hydrogen
absorption spectra of Kes79.

In radio W49B is a bright shell type SNR with radius $\sim$ 2 $\myarcmin\mbox{ }$, brightest along
the west and southeast edges. (Moffett \& Reynolds, 1994). In X-ray it is a center-brightened SNR and
the diffuse emission is extended all over the source (see \rosat\ HRI image).
Based on  the \asca\ data, Fujimoto et al. (1995) concluded that its spectrum has a thermal nature and
multiple plasma components are required to explain this spectrum. They also found that the iron ions
are mostly confined to the inner part.
Its distance is about 8 kpc (Radhakrishnan et al. 1972 ; Moffett \& Reynolds, 1994).

\section{DATA REDUCTION AND ANALYSIS}

\subsection{\underline{Kes79}}

Kes79 was observed by \asca\ from April 22 to April 23, 1995. Approximate exposure time is 37.1 ks
for SIS and 40 ks for GIS. SIS data are in 1-CCD mode. All data are screened using the standard
REV2 processing. After screening we obtain about 20000 events from SIS0 and about 16000 events 
from SIS1 (here we only analyse the SIS data).

The background-subtracted SIS spectrum of Kes79 is shown in Figure 1. The background spectrum is taken
from the Galactic Ridge observations that are near to Kes79. We can see the Fe-L "clump", Mg He$\alpha$, Si He$\alpha$
, Si He$\beta$ and S He$\alpha$ clearly in this spectrum. Then we consider to use a NEI model to fit its 
spectrum. The code that we adopt here is the SPEX code (Kaastra et al.\ 1993).
The fitting result is in Table 1. It is clearly shown in Table 1 that a single NEI component is
good enough to describe the spectrum of Kes79. The best-fit value of n$_{e}$t is much lower than
1.0$\times10^{12}$ cm$^{-3}$s, which implies that the plasma in Kes79 still doesn't reach 
equilibrium ionization. At that situation the emission of the lines tend to be much stronger
than in equilibrium. To show this we make the abundance of all the heavy elements in the best fit
(Table 1) as zero and keep the other parameters as those in the best fit. We call it "continuum" and
estimate that only 30\% of the emission comes from the so-called "continuum", so the remainder all comes
from the emission of the heavy elements. Then we make some narrow
band images of Kes79 to see the distribution of heavy elements in this source. The results are
shown in Figure 2. In this figure we can see that the emission of "continuum" is more center-brightened
than others. The images of (b) - (d) are all most bright in the S, SE and SW of Kes79, where
there may exists interaction between shocks and molecular clouds, but there also exists some difference
between these images. Though there indeed exist the difference of absorption between the N and
the S (about 1.65$\times10^{22}$ cm$^{-2}$, 1.95$\times10^{22}$ cm$^{-2}$ respectively
from the analysis of \rosat\ spectra), we find that this difference can't explain
the change of surface brightness from the N to the S. Due to its relative high energy,
the narrow band image of Si (plus continuum) is not sensitive to the change of
absorption. In Figure 2, We don't find clearly brightening in the N, even
in (c) and (d). So we may conclude that the heavy elements are richer in
the S, SE and SW of Kes79 and the N of Kes79 is intrinsically fainter than the other
parts.
Where do those heavy elements come from, those excavated from the molecular clouds or the ejecta stopped by
the molecular clouds? We need further observations.


\begin{figure}
\begin{minipage}{9cm}
\vspace{0.2cm}
\doublerulesep 1pt
 \centerline{\begin{tabular}{c|c}\hline\hline
   Parameter   &    Value$^{\rm\  a}$\\\hline
   N$_{\rm H} (10^{22}$ cm$^{-2})$  &  1.75$^{+0.07}_{-0.07}$ \\
   Temperature (keV)    &   0.88$^{+0.07}_{-0.07}$\\
   n$_{e}$t (10$^{10}$ cm$^{-3}$s)   & 5.9$^{+1.4}_{-1.0}$\\
   Ne  & 0.56$^{+0.43}_{-0.38}$      \\
   Mg  & 1.8$^{+0.3}_{-0.3}$      \\
   Si  & 1.2$^{+0.2}_{-0.1}$      \\
   S   & 1.1$^{+0.2}_{-0.2}$      \\
   Fe  & 1.8$^{+0.4}_{-0.4}$      \\
   $\chi^{2}$    &   131.6/123 (d.o.f)    \\ \hline\hline
 \end{tabular}}
\begin{flushleft}
\leftskip 10pt
\vspace{0.5cm}
$^{\rm a}$ Single-parameter 2.706~$\sigma$ errors\\
\end{flushleft}
\vspace{0.5cm}
\centerline{Table 1: Single NEI fit to the spectrum of Kes79}
\end{minipage}
\begin{minipage}{8.3cm}
\psfig{figure=fig1.ps,width=8.3cm,angle=270}
\centerline{Fig.1 : The spectrum of Kes79 and best fit model}
\end{minipage}
\end{figure}

\begin{figure}
\centerline{\psfig{figure=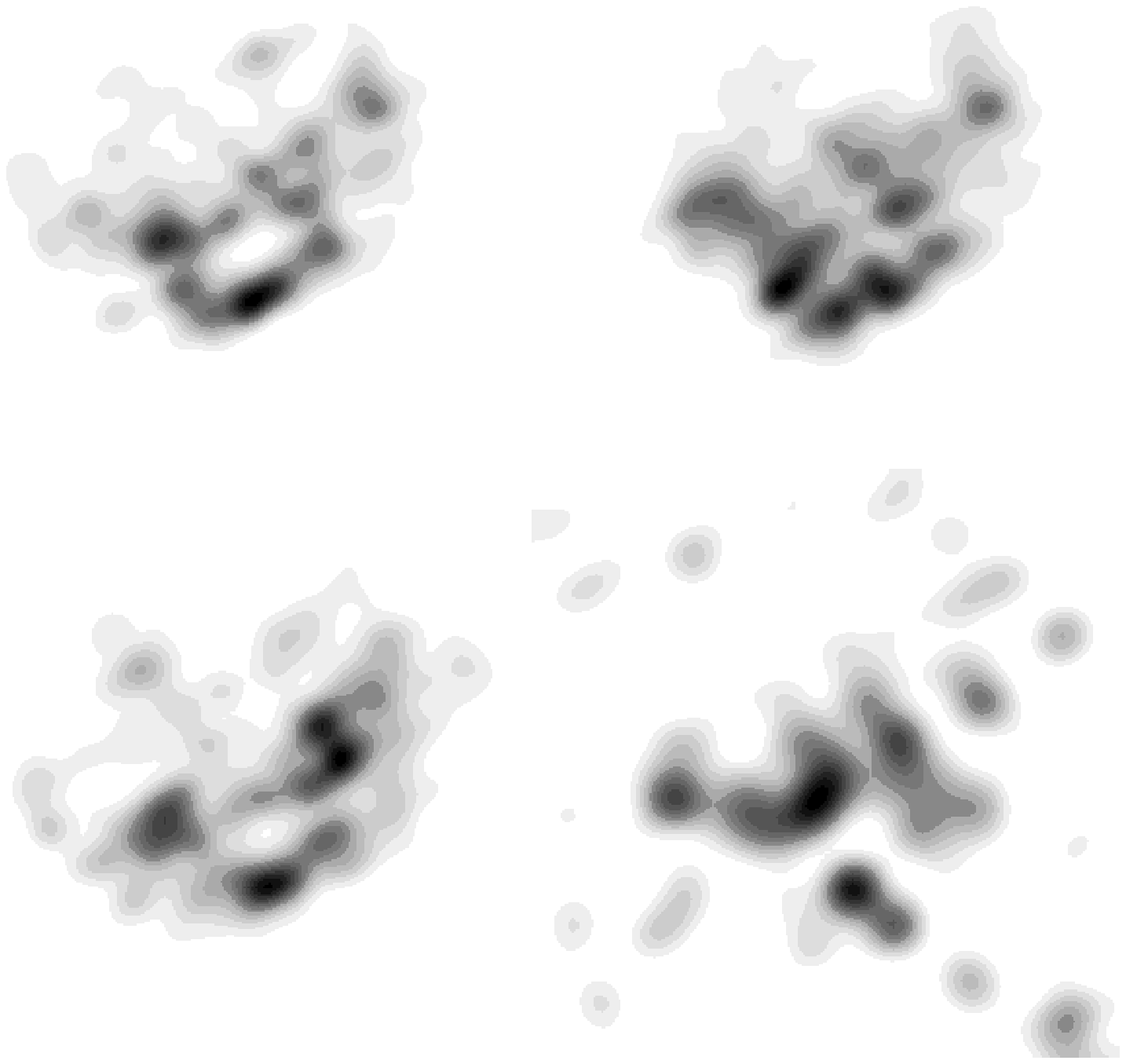,width=8cm}}
\vspace{2cm}
\centerline{Fig.2 : Narrow band images of Kes79 : (a) - (d) from the left to the right, the top to the bottom.
(a):}
\centerline{0.5-10keV; (b): 1.2-1.4keV, continuum + Mg He$\alpha$ lines; (c): 1.7-2.0keV, continuum
+ Si He$\alpha$ lines; (d):}
\centerline{2.6-10keV, mainly continuum. Each image covers the same region. They are
all exposure-corrected and}
\centerline{vignetting-corrected. (a)-(c) are deconvolved images. Due to the limited
photons, (d) is only smoothed.}
The size of each image is about $10\myarcmin\mbox{ }$.
\end{figure}

\subsection{\underline{W49B}}

W49B was observed by \asca\ several times. Here we mainly use the SIS data in April 24, 1993 (PV phase)
and use other data as a supplement (e.g. estimate the background). Approximate exposure time is 49.4 ks
for SIS. All data are screened using the standard REV2 processing. After screening we obtain about 
67000 events from SIS0 and about 52000 events from SIS1. The RDD effect should not be severe in the
PV phase but we still only use 2CCD mode data to analyse its spectrum.

First we obtained the spectrum of W49B (Figure 4). The emission lines of Si, S, Ar, Ca, Fe-K are all
clear in this spectrum. The emission lines of Mg, Ne and Fe-L can't be seen due to the high
absorption. We notice that there seems to exist an emission line at about 5.7 keV. Based on the data from
the best calibrated chip (S0 chip1), the centroid of this line is 5.71$^{+0.06}_{-0.08}$ keV.
The
expected line centroid of the Cr He$\alpha$ lines is : 5.63-5.68 keV (Shirai et al. 1993).
Considering the uncertainty of the calibration,
these two value are consistent to each other. But more observations with
more exposure time are needed to clarify the nature of this possible line. From previous work (e.g. Fujimoto et al. 1995)
We know that the spectrum of W49B is too complicated to get satisfactory fits, but
the emission lines shown in the spectrum provide a good opportunity to perform plasma diagnostics.
The diagnostics are : the ratio of Ly$\alpha$ / He$\alpha$ and He$\beta$ (or He3p(+4p)) / He$\alpha$;
the energy centroid of He$\alpha$ lines. We list these value in Table.2 and the results
in Figure 3. Our results have some difference from the results of Fujimoto et al. (1995) mainly
because of different plasma codes we applied.
In Figure 3 we can see that iron, argon and calcium are all very hot, at least hotter than 2 keV if we admit
that the plasma in W49B is now in NEI situation because of its young age (Fujimoto et al.1995).
And we can also see in Figure 3 that at least two components of Silicon and Sulfur are needed.
Now we begin to try spectral fitting. A single NEI component fit is bad, so we consider to try a two component NEI fit. As implied by Figure 3 and in order to
prevent too many free parameters, we make the abundances of Fe, Ca, Ar in the second component
 zero and fix N$_{\rm H}$ at 4$\times10^{22}$ cm$^{-2}$. The results are shown in Table 3. The
first component is a hard component and its overabundance implies that it may come from the shocked 
ejecta. The second soft component it may come from the diffuse ejecta. We don't find
evidence for a hot blast wave, even in the GIS spectrum.

\begin{figure}
\hspace*{0.2cm}
\vspace{-0.5cm}
\begin{minipage}{8cm}
\psfig{figure=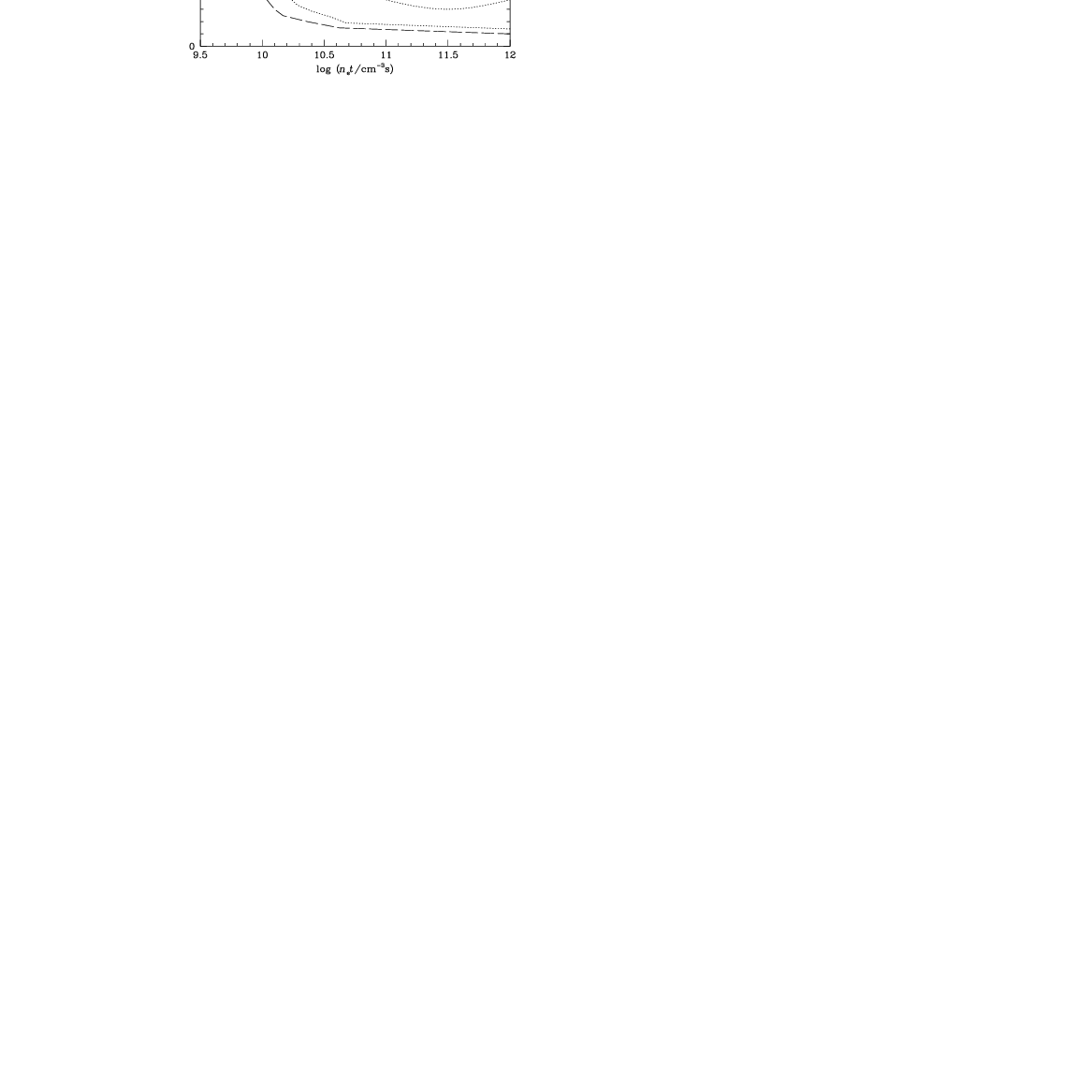,width=8cm}
\vspace{1cm}
\centerline{Fig.3 : The derived ranges for Si, S, Ar, Ca and}
\centerline{Fe by line ratios in the n$_{e}$t-T plane based on the}
\centerline{SPEX code. S1 means the region determined by}
\centerline{the ratio He$\beta$ (or He3p(+4p) ) / He$\alpha$; S2 means }
\centerline{the region  determined by the  ratio Ly$\alpha$ / He$\alpha$. }
Si1 and Si2 are similar as the above mentioned.
\end{minipage}
\hspace*{0.1cm}
\begin{minipage}{8.5cm}
\vspace*{0.8cm}
\vspace{-0.8cm}
\psfig{figure=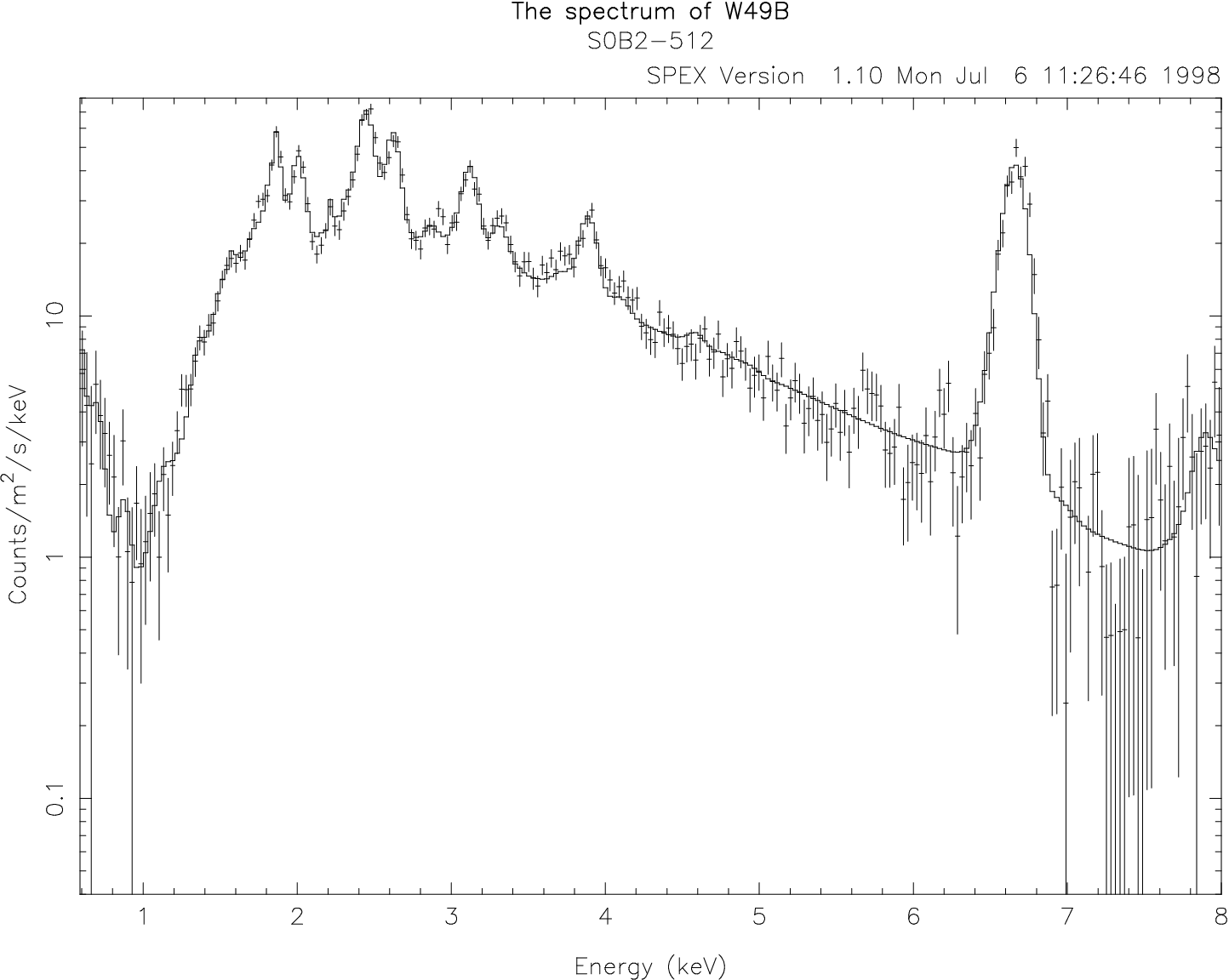,width=8cm,angle=270}
\centerline{Fig.4 : The spectrum of W49B and two-NEI fit}
\end{minipage}
\end{figure}

\vspace{3cm}
\doublerulesep 1pt
 \centerline{\begin{tabular}{c|c}\hline\hline
   Ratio    &    Value$^{\rm\  a}$\\\hline\hline
   Si He(3p+4p)/Si He$\alpha$  &  0.081$\pm$0.050 \\
   Si Ly$\alpha$/Si He$\alpha$  &   0.497$\pm$0.068\\
   S He(3p)/S He$\alpha$ & 0.037$\pm$0.024\\
   S Ly$\alpha$/S He$\alpha$  & 0.520$\pm$0.053\\
   Ar Ly$\alpha$/Ar He$\alpha$ & 0.376$\pm$0.079\\
   Ca Ly$\alpha$/Ca He$\alpha$ & 0.204$\pm$0.080\\
   Fe Ly$\alpha$/Fe He$\alpha$ & $<$ 0.020 \\\hline
   Energy centroid  &  Value$^{\rm\  a}$\\\hline\hline
   Fe$^{\rm\  b}$ (keV) & 6.660$^{+0.007}_{-0.006}$\\ \hline\hline
 \end{tabular}}
\begin{flushleft}
\leftskip 150pt
$^{\rm a}$ Single-parameter 2.706~$\sigma$ errors\\
$^{\rm b}$ Using the best calibrated chip SIS0 chip1.\\
\end{flushleft}
Table 2: Some diagnostic values of W49B (based on the fit
using a bremsstrahlung continuum and some Gaussians)

\section{DISCUSSION}

\subsection{\underline{Kes79}}
In the above analysis we see that the heavy elements are inhomogeneously distributed in Kes79.
Seward \& Velusamy once concluded that the variation of the spectra around this source is
little by analysing the \rosat\ data, when combining with the single NEI component nature of
its spectrum, it seems that the plasma in this source is in an isothermal condition, which may
come from the thermal conduction. There still remains a question. Where do these heavy elements
come from? Though their abundance is only a little higher than solar, from Figure 2 we
see that the emission of these heavy elements is mainly concentrated at the S, SW and SE of Kes79,
where there may exist the interaction of the shock with molecular clouds. And we see that the emission of the 
continuum is center-brighted. So it seems that the heavy elements are rich in that region.
But as it now stands, we can't distinguish whether the heavy elements come from the ejecta stopped
by the dense molecular clouds, or from the molecular clouds, or from both of them. Further observation
of higher spatial and spectral resolution are needed.

\subsection{\underline{W49B}}
As mentioned in the above, the ejecta in W49B may be still hot, even hotter than the blast wave. This
case is somewhat similar to the case of Tycho (Hwang et al. 1998), where the Fe ejecta have a higher
temperature than the Si, S ejecta (maybe also Ar, Ca ejecta). It is against the result of
Chevalier (1982) under the assumption of the power-law ejecta profiles, which would imply higher
density and lower temperature at inner radii. Recently Dwarkadas \& Chevalier (1998) obtained a 
relatively flat temperature profile under the assumption of the exponential ejecta density profile
for Type Ia SN. And the presence of a small amount of circumstellar matter may increase
the temperature contrast. But now, before getting more data with better spatial and spectral
resolution, it is hard to make a definite conclusion. And there is still another possibility
that the blast wave is very hot and has quite a small emission messure, so now we can't
seperate this component from the spatially integrated spectrum. As for the Cr line, it also needs
further investigation due to its low signal-to-noise in \asca\ observation now.

\vspace{0.5cm}
\doublerulesep 1pt
\centerline{\begin{tabular}{c|c|c|c|c|c|c|c}\hline\hline
 Parameter  & Value & - & - & - & - & - & -\\\hline
 EM$_{1}$ (10$^{58}$ cm$^{-3}$) & 7.0 & 6.0 & 5.0 & 4.0 & 2.0 & 1.0 & 0.5\\
 log(n$_{e}$t$_{1}$/cm$^{-3}s$)& 11.92 & 11.93 &11.89&11.87&11.80&11.71&11.76\\
 kT$_{1}$ (keV) & 1.98 & 2.06 & 2.12 & 2.18  & 2.34 & 2.52 & 2.41\\
 Si$_{1}$  & 2.2 & 2.6 & 3.0 & 3.8 & 7.2 & 13. & 28.\\
 S$_{1}$  & 3.2 & 3.7 & 4.4 & 5.3 & 9.9 & 18. & 38.\\
 Ar$_{1}$ & 3.4 &4.1 & 4.8 & 6.0 & 12. & 22. & 46.\\
 Ca$_{1}$ & 4.6 & 5.3 & 6.3 & 7.8 & 15. & 28. & 58.\\
 Fe$_{1}$  & 3.1 & 3.5 & 4.0 & 4.7 & 8.4 & 15. & 32.\\
 M$_{1}^{\rm c}$ (M$_{\odot}$)& 6.7 & 6.2 & 5.6 & 5.0 & 3.6 & 2.5 & 2.0\\
 EM$_{2}$ (10$^{58}$cm$^{-3}$) & 5.5 & 7.4 & 8.5 & 9.4 & 11.3 & 12.1 & 12.4\\
 log(n$_{e}$t$_{2}$/cm$^{-3}s$) & 10.15 &10.13 & 10.13 &10.13&10.14&10.15&10.13\\
 kT$_{2}$ (keV) & 0.97 & 1.17 & 1.23 & 1.31 & 1.44 & 1.52 & 1.57\\
 Si$_{2}$ & 1.7 & 0.95 & 0.78 & 0.64 & 0.47 & 0.41 & 0.39\\
 S$_{2}$  & 0.49 & 0.52 & 0.47 & 0.46 & 0.43 & 0.43 & 0.37\\
 M$_{2}^{\rm c}$ (M$_{\odot}$) & 13. & 15.& 17. & 18. & 20. & 21. & 21.\\
 $\chi^{2}$ (236 d.o.f)&391.3 &379.6 &376.3 &373.1 &369.5 &367.6&368.8\\ \hline\hline
\end{tabular}}
\vspace{-0.2cm}
\begin{flushleft}
\leftskip 32pt
$^{\rm a}$ N$_{H}$ is fixed to 4.0$\times10^{22}$ cm$^{-2}$.\\
$^{\rm b}$ Ar, Ca, Fe in component 2 are fixed to 0.\\
$^{\rm c}$ The mass of heavy elements are all included and take f$_{1}$ as 0.05, f$_{2}$ as 0.25.\\
note: n$_{e}$/n$_{H}$ = 1.20 - 1.27 according to the fitting results.\\
\end{flushleft}
\centerline{Table 3: Spectral fitting results with two NEI components for the spectrum of W49B}
 
\section{REFERENCES}
\vspace{-5mm}
\begin{itemize}
\setlength{\itemindent}{-8mm}
\setlength{\itemsep}{-1mm}

\item[]
Chevalier R. A. {\bf ApJ}, 258, 790 (1982)

\item[]
Dwarkadas V. V., Chevalier R. A., {\bf ApJ}, 497, 807 (1998)

\item[]
Frail D.A., Clifton T.R. {\bf ApJ}, 336, 854 (1989)

\item[] 
Fujimoto R. et al. {\bf PASJ}, 47, L31 (1995)

\item[]
Green A.J., Frail D.A., Goss W.M., Otrupcek R. {\bf AJ}, 114, 2269 (1997)

\item[]
Green D.A., Dewdney P.E. {\bf MNRAS}, 254, 686 (1992)

\item[]
Hwang U., Hughes J.P., Petre R. {\bf ApJ}, 497, (1998), astro-ph/9712241

\item[]
Kaastra, J. S., Mewe, R., Nieuwenhuijzen, H. 1996, in The 11th coll.\ on UV and X-ray Spectr.\ of Astroph.\ and Laboratory Plasmas, Watanabe, T.(ed.), p.411

\item[]
Moffett D.A., Reynolds S.P. {\bf ApJ}, 437, 705 (1994)

\item[]
Radhakrishnan V., Goss W.M., Murray J.D., Brooks J.W. {\bf ApJS}, 24, 49 (1972)

\item[]
Scoville N.Z., Yun M.S., Clemens D.P., Sanders D.B., Waller W.H. {\bf ApJS}, 63, 821 (1987)
 
\item[]
Seward F.D., Velusamy T. {\bf ApJ}, 439, 715 (1995)

\item[]
Shirai T., Nakai Y., Nakagaki T., Sugar J. Wiese W.L. {\bf J. Phys. Chem. Ref. Data}, 22, 1281 (1993)

\item[]
Velusamy T., Becker R.H., Seward F.D. {\bf AJ}, 102, 676 (1991)

\end{itemize}

\end{document}